\documentclass{article}

\usepackage[dblblindworkshop, final]{neurips_2026}
\workshoptitle{EvoRobust}

\usepackage[utf8]{inputenc}
\usepackage[T1]{fontenc}
\usepackage{hyperref}
\hypersetup{
  pdftitle    = {Evolving Decoding Algorithms for Quantum Error Correction},
  pdfauthor   = {Eduardo Dueñez, Edgar Said Hernandez Sanchez, Edgar A. Duéñez-Guzmán, Suzanne Sadedin},
  pdfsubject  = {EvoRobust workshop, NeurIPS 2026},
  pdfkeywords = {quality-diversity search, MAP-Elites, genetic programming, program synthesis, domain-specific language, algorithm discovery, multi-objective optimization, robustness evaluation, quantum error correction, quantum decoder design},
}
\usepackage{url}
\usepackage{booktabs}
\usepackage{amsfonts}
\usepackage{amsmath}
\usepackage{nicefrac}
\usepackage{microtype}
\usepackage{xcolor}
\usepackage{graphicx}
\usepackage{enumitem}
\usepackage{array}

\newcommand{\SepSplit}{\texttt{s20260904}}
\newcommand{\SepDigest}{\texttt{e49376e681002aba}}
\newcommand{\SepShots}{200\,000}

\newcommand{\SepCThreeTimeXPTwoD}{0.64}
\newcommand{\SepCTwelveLer}{0.1538}
\newcommand{\SepPTwoDLer}{0.05454}
\newcommand{\SepPTwoDMs}{8.82}
\newcommand{\SepLdpcOsdZeroLer}{0.04392}
\newcommand{\SepLdpcOsdZeroTimeXPTwoD}{103}
\newcommand{\SepRelayFiveMs}{190.0}
\newcommand{\SepRelayNineLer}{0.02795}
\newcommand{\SepRelayNineMs}{270.6}
\newcommand{\SepLdpcOsdZeroMsExact}{903.75}
\newcommand{\SepCThreeLdpcOverThis}{160}
\newcommand{\SepCThreeZVsPTwoD}{+31.8}
\newcommand{\SepCThreeZVsLdpcOsdZero}{+5.0}
\newcommand{\SepCThreeErrCutVsPTwoD}{23\%}
\newcommand{\SepCThreeErrCutVsLdpcOsdZero}{4.8\%}
\newcommand{\SepCFourLdpcOverThis}{44}
\newcommand{\SepCFourZVsLdpcOsdZero}{+6.9}
\newcommand{\SepCTenLdpcOverThis}{31}
\newcommand{\SepCTenZVsPTwoD}{+33.6}
\newcommand{\SepCTenZVsLdpcOsdZero}{+8.9}
\newcommand{\SepCTenZVsAc}{\ensuremath{-}18.9}
\newcommand{\SepCTenErrCutVsPTwoD}{25\%}
\newcommand{\SepLdpcOsdZeroZVsPTwoD}{+23.6}
\newcommand{\SepRelayFiveZVsAc}{+10.2}
\newcommand{\SepAcTimeXCThreeWord}{six}
\newcommand{\SepRelayFiveTimeXCThreeWord}{thirty-four}
\newcommand{\KnobDampedLerPct}{5.15\%}
\newcommand{\KnobUndampedLerPct}{47.66\%}
\newcommand{\KnobLerRatio}{9.3}
\newcommand{\KnobEdgesRatio}{1.04}
\newcommand{\KnobLiftedLerPct}{47.4\%}
\newcommand{\KnobLiftedMs}{96.0}
\newcommand{\OctDampSplit}{\texttt{s20261004}}
\newcommand{\OctDampShots}{200\,000}
\newcommand{\OctDampCpu}{Apple M3 Ultra}
\newcommand{\OctDampUnitStart}{8\,909.5}
\newcommand{\OctDampUnitEnd}{8\,920.0}
\newcommand{\OctDampDrift}{0.12\%}
\newcommand{\OctDampBlockShots}{256}
\newcommand{\OctDampBurnIn}{32}
\newcommand{\OctDampPTwoLer}{0.04355}
\newcommand{\OctDampPTwoMs}{365.08}
\newcommand{\OctDampPTwoSettings}{product-sum, serial, 30 it., undamped; OSD-0}
\newcommand{\OctDampPTwoDLer}{0.05525}
\newcommand{\OctDampPTwoDMs}{8.91}
\newcommand{\OctDampPTwoDSettings}{min-sum 0.625, flooding, 10 it., damping 0.3; OSD-0}
\newcommand{\OctDampPSevenLer}{0.05080}
\newcommand{\OctDampPSevenMs}{364.24}
\newcommand{\OctDampPSevenDLer}{0.07241}
\newcommand{\OctDampPSevenDMs}{16.25}
\newcommand{\OctDampPSevenDSettings}{min-sum 0.625, flooding, 30 it., damping 0.3; LSD-0 to validity}
\newcommand{\OctDampPEightLer}{0.05048}
\newcommand{\OctDampPEightMs}{364.17}
\newcommand{\OctDampPEightDLer}{0.07206}
\newcommand{\OctDampPEightDMs}{16.26}
\newcommand{\OctDampPThreeLer}{0.03841}
\newcommand{\OctDampPThreeMs}{372.35}
\newcommand{\OctDampPThreeOursSettings}{as P2-d; sweep weight \ensuremath{\le} 2 over the top 100 free columns}
\newcommand{\OctDampPThreeDLer}{0.04224}
\newcommand{\OctDampPThreeDMs}{21.71}
\newcommand{\OctDampPThreeDSettings}{as P2-d; OSD-CS order 60 (P3's back end)}
\newcommand{\OctDampOsdZeroSpeedup}{41.0}
\newcommand{\OctDampOsdZeroZ}{+25.8}
\newcommand{\OctDampLsdZeroSpeedup}{22.4}
\newcommand{\OctDampLsdZeroZ}{+42.7}
\newcommand{\OctDampLsdMuSpeedup}{22.4}
\newcommand{\OctDampLsdMuZ}{+42.7}
\newcommand{\OctDampOsdCsFrontSpeedup}{17.1}
\newcommand{\OctDampOsdCsFrontZ}{+9.2}
\newcommand{\OctDampZMin}{+9.2}
\newcommand{\OctDampZMax}{+42.7}
\newcommand{\OctDampLerRatioMin}{1.10}
\newcommand{\OctDampLerRatioMax}{1.43}
\newcommand{\OctDampPThreeOursTimeXPThreeD}{3.4}
\newcommand{\OctAdjSplit}{\texttt{s20261004}}
\newcommand{\OctAdjDigest}{\texttt{0c07de8e383fe98d}}
\newcommand{\OctAdjShots}{200\,000}
\newcommand{\OctAdjMargin}{1.4}
\newcommand{\OctAdjSubjects}{sixteen}
\newcommand{\OctAdjRefusals}{zero}
\newcommand{\OctAdjProxyOverTimeMin}{1.01}
\newcommand{\OctAdjProxyOverTimeMax}{1.26}
\newcommand{\OctAdjUnitNOne}{14\,946}
\newcommand{\OctAdjUnitPanel}{13\,618}
\newcommand{\OctAdjDrift}{0.94\%}
\newcommand{\OctAdjPTwoDInNOneUnits}{1.016}
\newcommand{\OctAdjMachine}{Intel Xeon Platinum 8581C}
\newcommand{\OctAdjPrimaryProxy}{thirteen}
\newcommand{\OctAdjPrimaryTime}{twelve}
\newcommand{\OctAdjReachTime}{zero}
\newcommand{\OctAdjStretchTime}{four}
\newcommand{\OctAdjStretchWinners}{C03, C07, C08, C09}
\newcommand{\OctAdjCThreeLer}{0.04173}
\newcommand{\OctAdjCThreeCostXPTwoD}{0.58}
\newcommand{\OctAdjCThreeTimeXPTwoD}{0.565}
\newcommand{\OctAdjCThreeZVsPTen}{\ensuremath{-}19.6}
\newcommand{\OctAdjCThreeZVsPTwoD}{+33.4}
\newcommand{\OctAdjCFourLer}{0.04065}
\newcommand{\OctAdjCFourCostXPTwoD}{2.0}
\newcommand{\OctAdjCFourTimeXPTwoD}{1.882}
\newcommand{\OctAdjCFourZVsPTen}{\ensuremath{-}20.6}
\newcommand{\OctAdjCFourZVsPTwoD}{+35.8}
\newcommand{\OctAdjCFiveLer}{0.06626}
\newcommand{\OctAdjCFiveCostXPTwoD}{2.4}
\newcommand{\OctAdjCFiveTimeXPTwoD}{1.887}
\newcommand{\OctAdjCFiveZVsPTen}{\ensuremath{-}66.1}
\newcommand{\OctAdjCFiveZVsPTwoD}{\ensuremath{-}22.4}
\newcommand{\OctAdjCSixLer}{0.05584}
\newcommand{\OctAdjCSixCostXPTwoD}{1.0}
\newcommand{\OctAdjCSixTimeXPTwoD}{0.995}
\newcommand{\OctAdjCSixZVsPTen}{\ensuremath{-}49.4}
\newcommand{\OctAdjCSixZVsPTwoD}{\ensuremath{-}1.7}
\newcommand{\OctAdjCSevenLer}{0.04694}
\newcommand{\OctAdjCSevenCostXPTwoD}{0.46}
\newcommand{\OctAdjCSevenTimeXPTwoD}{0.456}
\newcommand{\OctAdjCSevenZVsPTen}{\ensuremath{-}31.9}
\newcommand{\OctAdjCSevenZVsPTwoD}{+23.3}
\newcommand{\OctAdjCEightLer}{0.04159}
\newcommand{\OctAdjCEightCostXPTwoD}{0.57}
\newcommand{\OctAdjCEightTimeXPTwoD}{0.534}
\newcommand{\OctAdjCEightZVsPTen}{\ensuremath{-}19.6}
\newcommand{\OctAdjCEightZVsPTwoD}{+33.1}
\newcommand{\OctAdjCNineLer}{0.04315}
\newcommand{\OctAdjCNineCostXPTwoD}{0.54}
\newcommand{\OctAdjCNineTimeXPTwoD}{0.537}
\newcommand{\OctAdjCNineZVsPTen}{\ensuremath{-}22.5}
\newcommand{\OctAdjCNineZVsPTwoD}{+30.8}
\newcommand{\OctAdjCTenLer}{0.04036}
\newcommand{\OctAdjCTenCostXPTwoD}{2.7}
\newcommand{\OctAdjCTenTimeXPTwoD}{2.389}
\newcommand{\OctAdjCTenZVsPTen}{\ensuremath{-}22.0}
\newcommand{\OctAdjCTenZVsPTwoD}{+35.9}
\newcommand{\OctAdjCElevenLer}{0.06619}
\newcommand{\OctAdjCElevenCostXPTwoD}{2.4}
\newcommand{\OctAdjCElevenTimeXPTwoD}{1.955}
\newcommand{\OctAdjCElevenZVsPTen}{\ensuremath{-}66.0}
\newcommand{\OctAdjCElevenZVsPTwoD}{\ensuremath{-}22.2}
\newcommand{\OctAdjCTwelveLer}{0.1551}
\newcommand{\OctAdjCTwelveCostXPTwoD}{1.8}
\newcommand{\OctAdjCTwelveTimeXPTwoD}{1.658}
\newcommand{\OctAdjCTwelveZVsPTen}{\ensuremath{-}146.7}
\newcommand{\OctAdjCTwelveZVsPTwoD}{\ensuremath{-}126.9}
\newcommand{\OctAdjETwentyFiveLer}{0.03572}
\newcommand{\OctAdjETwentyFiveCostXPTwoD}{1.1}
\newcommand{\OctAdjETwentyFiveTimeXPTwoD}{1.011}
\newcommand{\OctAdjETwentyFiveZVsPTen}{\ensuremath{-}4.1}
\newcommand{\OctAdjETwentyFiveZVsPTwoD}{+46.6}
\newcommand{\OctAdjETwentySevenLer}{0.03206}
\newcommand{\OctAdjETwentySevenCostXPTwoD}{2.4}
\newcommand{\OctAdjETwentySevenTimeXPTwoD}{2.145}
\newcommand{\OctAdjETwentySevenZVsPTen}{+9.4}
\newcommand{\OctAdjETwentySevenZVsPTwoD}{+54.7}
\newcommand{\OctAdjETwentySevenTimeXAc}{0.80}
\newcommand{\OctAdjPTenLer}{0.03435}
\newcommand{\OctAdjPTenCostXPTwoD}{3.0}
\newcommand{\OctAdjPTenTimeXPTwoD}{2.67}
\newcommand{\OctAdjPTwoLer}{0.04355}
\newcommand{\OctAdjPTwoCostXPTwoD}{25.1}
\newcommand{\OctAdjPTwoDLer}{0.05525}
\newcommand{\OctAdjPTwoDCostXPTwoD}{1.0}
\newcommand{\OctAdjPTwoDTimeXPTwoD}{1.00}
\newcommand{\OctAdjPThreeLer}{0.03841}
\newcommand{\OctAdjPThreeCostXPTwoD}{26.0}
\newcommand{\OctAdjPFiveLer}{0.02856}
\newcommand{\OctAdjPFiveCostXPTwoD}{40.0}
\newcommand{\OctAdjPFiveTimeXPTwoD}{35.57}
\newcommand{\OctAdjETwoLer}{0.03020}
\newcommand{\OctAdjETwoCostXPTwoD}{2.9}
\newcommand{\OctAdjETwoRaisedShots}{42}
\newcommand{\OctRelaySplit}{\texttt{s20261004}}
\newcommand{\OctRelayShots}{20\,000}
\newcommand{\OctRelayUnitMs}{8.77}
\newcommand{\OctRelayFiveMs}{172.8}
\newcommand{\OctRelayFiveTimeXPTwoD}{19.7}
\newcommand{\OctRelayNineMs}{272.0}
\newcommand{\OctRelayNineTimeXPTwoD}{31.0}
\newcommand{\GridPoints}{twenty-two}
\newcommand{\GridPointsN}{22}
\newcommand{\GridAcFront}{seventeen}
\newcommand{\GridAcFrontN}{17}
\newcommand{\GridCThreeFrontN}{22}
\newcommand{\GridCFourFront}{thirteen}
\newcommand{\GridCFourFrontN}{13}
\newcommand{\GridCSevenFrontN}{0}
\newcommand{\GridCEightFrontN}{1}
\newcommand{\GridCNineFrontN}{22}
\newcommand{\GridCTenFront}{eleven}
\newcommand{\GridCTenFrontN}{11}
\newcommand{\GridPTwoDFrontN}{0}
\newcommand{\GridLdpcOsdZeroFrontN}{0}
\newcommand{\GridLdpcOsdCsFrontN}{1}
\newcommand{\GridRelayFiveFront}{ten}
\newcommand{\GridRelayFiveFrontN}{10}
\newcommand{\GridRelayNineFront}{sixteen}
\newcommand{\GridRelayNineFrontN}{16}
\newcommand{\GridShotsMin}{3\,500}
\newcommand{\GridShotsMax}{10\,216\,500}
\newcommand{\GridNOneHundredFortyFourShots}{24\,500}
\newcommand{\GridNTwoHundredEightyEightShots}{40\,000}
\newcommand{\GridPOneShots}{10\,216\,500}
\newcommand{\GridREighteenShots}{6\,500}
\newcommand{\GridPSixShots}{3\,500}
\newcommand{\GridNTwoHundredEightyEightLdpcOsdCsShots}{4\,000}
\newcommand{\GridNOneHundredFortyFourAcErrors}{431}
\newcommand{\GridNOneHundredFortyFourAcMs}{211.6}
\newcommand{\GridNOneHundredFortyFourCThreeErrors}{310}
\newcommand{\GridNOneHundredFortyFourCThreeMs}{29.9}
\newcommand{\GridNOneHundredFortyFourPTwoDErrors}{816}
\newcommand{\GridNOneHundredFortyFourPTwoDMs}{79.3}
\newcommand{\GridNOneHundredFortyFourRelayFiveErrors}{448}
\newcommand{\GridNOneHundredFortyFourRelayFiveMs}{2\,553}
\newcommand{\GridPOneAcErrors}{718}
\newcommand{\GridPOneAcMs}{7.23}
\newcommand{\GridPOneCThreeErrors}{760}
\newcommand{\GridPOneCThreeMs}{2.30}
\newcommand{\GridPOneCEightErrors}{633}
\newcommand{\GridPOneCEightMs}{3.83}
\newcommand{\GridPOnePTwoDErrors}{996}
\newcommand{\GridPOnePTwoDMs}{3.22}
\newcommand{\GridPOneRelayFiveErrors}{915}
\newcommand{\GridPOneRelayFiveMs}{22.7}
\newcommand{\SepAblShots}{20\,000}
\newcommand{\SepAblCThreeLer}{0.04135}
\newcommand{\SepAblCThreeMs}{5.57}
\newcommand{\SepAblCNineLer}{0.04415}
\newcommand{\SepAblCNineMs}{5.43}
\newcommand{\SepAblCThreeProductSumLer}{0.03680}
\newcommand{\SepAblCThreeProductSumMs}{10.7}
\newcommand{\SepAblCThreeProductSumZ}{+3.8}
\newcommand{\SepAblCThreeProductSumTimeXBase}{1.9}
\newcommand{\SepAblCThreeKappaLer}{0.04175}
\newcommand{\SepAblCThreeKappaMs}{6.11}
\newcommand{\SepAblCThreeKappaZ}{\ensuremath{-}0.8}
\newcommand{\SepAblCThreeKappaTimeXBase}{1.1}
\newcommand{\SepAblCThreeWeightTwoLer}{0.04165}
\newcommand{\SepAblCThreeWeightTwoMs}{5.67}
\newcommand{\SepAblCThreeWeightTwoZ}{\ensuremath{-}0.3}
\newcommand{\SepAblCThreeWeightTwoTimeXBase}{1.0}
\newcommand{\SepAblCThreeAllThreeLer}{0.03660}
\newcommand{\SepAblCThreeAllThreeMs}{11.2}
\newcommand{\SepAblCThreeAllThreeZ}{+4.0}
\newcommand{\SepAblCThreeAllThreeTimeXBase}{2.0}
\newcommand{\SepAblCNineEarlyExitLer}{0.04415}
\newcommand{\SepAblCNineEarlyExitMs}{5.40}
\newcommand{\SepAblCNineEarlyExitZ}{0.0}
\newcommand{\SepAblCNineEarlyExitTimeXBase}{0.99}
\newcommand{\SepAblCNineProductSumLer}{0.04075}
\newcommand{\SepAblCNineProductSumMs}{10.1}
\newcommand{\SepAblCNineProductSumZ}{+3.0}
\newcommand{\SepAblCNineProductSumTimeXBase}{1.9}
\newcommand{\SepAblCNineKappaLer}{0.04535}
\newcommand{\SepAblCNineKappaMs}{6.30}
\newcommand{\SepAblCNineKappaZ}{\ensuremath{-}3.8}
\newcommand{\SepAblCNineKappaTimeXBase}{1.2}
\newcommand{\SepAblAcLer}{0.03475}
\newcommand{\SepAblAcMs}{31.3}
\newcommand{\SepAblAcZ}{+5.6}
\newcommand{\SepAblAcTimeXBase}{5.6}
\newcommand{\SepAblPTwoDLer}{0.05575}
\newcommand{\SepAblPTwoDMs}{8.90}
\newcommand{\SepAblPTwoDZ}{\ensuremath{-}11.2}
\newcommand{\SepAblPTwoDTimeXBase}{1.6}
\newcommand{\NovSubjectsListed}{seven}
\newcommand{\NovConfirmShots}{32\,768}
\newcommand{\NovCThreeArm}{seeded, work proxy}
\newcommand{\NovCThreeNearest}{P10 (BP+AC)}
\newcommand{\NovCThreeDistPruned}{0}
\newcommand{\NovCThreeCallsVerbatimN}{5}
\newcommand{\NovCThreeCallsPruned}{four}
\newcommand{\NovCThreeCallsPrunedN}{4}
\newcommand{\NovCThreeVerdictFull}{retuning of BP+AC (min-sum inside BP+AC)}
\newcommand{\NovCFourArm}{unseeded, work proxy}
\newcommand{\NovCFourNearest}{P3 (tool) / P2 (pruned)}
\newcommand{\NovCFourDistPruned}{2}
\newcommand{\NovCFourCallsVerbatimN}{15}
\newcommand{\NovCFourCallsPrunedN}{5}
\newcommand{\NovCFourVerdictFull}{recomposition of published stages (P4's decimation + P2's OSD-0); functionally P2 retuned}
\newcommand{\NovCSevenArm}{seeded, wall-clock}
\newcommand{\NovCSevenNearest}{P10 (BP+AC)}
\newcommand{\NovCSevenDistPruned}{5}
\newcommand{\NovCSevenCallsVerbatimN}{24}
\newcommand{\NovCSevenCallsPrunedN}{9}
\newcommand{\NovCSevenVerdictFull}{recomposition: BP+AC + cut + frontier/reseed stage (panel ops)}
\newcommand{\NovCEightArm}{seeded, wall-clock}
\newcommand{\NovCEightNearest}{P10 (BP+AC)}
\newcommand{\NovCEightDistPruned}{1}
\newcommand{\NovCEightCallsVerbatimN}{17}
\newcommand{\NovCEightCallsPrunedN}{6}
\newcommand{\NovCEightVerdictFull}{recomposition: BP+AC + P12's degeneracy cut; mostly retuning}
\newcommand{\NovCNineArm}{seeded, wall-clock}
\newcommand{\NovCNineNearest}{P10 (BP+AC)}
\newcommand{\NovCNineDistPruned}{0}
\newcommand{\NovCNineCallsVerbatimN}{4}
\newcommand{\NovCNineCallsPrunedN}{4}
\newcommand{\NovCNineVerdictFull}{retuning of BP+AC (min-sum inside BP+AC)}
\newcommand{\NovCTenArm}{unseeded, wall-clock}
\newcommand{\NovCTenNearest}{P2 (BP+OSD-0)}
\newcommand{\NovCTenDistPruned}{0}
\newcommand{\NovCTenCallsVerbatimN}{33}
\newcommand{\NovCTenCallsPrunedN}{5}
\newcommand{\NovCTenVerdictFull}{retuning of P2}
\newcommand{\NovETwentySevenArm}{seeded, work proxy}
\newcommand{\NovETwentySevenNearest}{P10 (BP+AC)}
\newcommand{\NovETwentySevenDistPruned}{0}
\newcommand{\NovETwentySevenCallsVerbatimN}{8}
\newcommand{\NovETwentySevenCallsPrunedN}{4}
\newcommand{\NovETwentySevenVerdictFull}{retuning of BP+AC}
\newcommand{\NovETwentySevenSweeps}{31}
\newcommand{\NovAcSweeps}{45}
\newcommand{\NovETwentySevenGamma}{0.412}
\newcommand{\NovAcGamma}{0.3}
\newcommand{\NovETwentySevenKappa}{0.0327}
\newcommand{\NovAcKappa}{0.1}
\newcommand{\NovETwentySevenErrorsBelowAc}{58}
\newcommand{\NovETwentySevenErrorsFromGamma}{57}
\newcommand{\NovETwentySevenCostXAc}{0.81}
\newcommand{\NovWordingCRows}{descendants of Ambiguity Clustering that keep its shape and differ in the settings}
\newcommand{\HypDslOps}{38}
\newcommand{\HypWStar}{0.0071}
\newcommand{\HypWorkCap}{60}
\newcommand{\HypBudgetSet}{\texttt{n1.m5b}}
\newcommand{\HypBudgetSweeps}{2\,000}
\newcommand{\HypBudgetOpCalls}{256}
\newcommand{\HypBudgetCandidates}{10\,100}
\newcommand{\HypBudgetColumnOps}{982\,256}
\newcommand{\HypTournament}{three}
\newcommand{\HypPrograms}{64}
\newcommand{\HypTrainShots}{1\,024}
\newcommand{\HypWorkers}{120}
\newcommand{\HypSeedMembers}{eighteen}
\newcommand{\HypRuns}{twelve}
\newcommand{\HypReplicates}{three}
\newcommand{\HypGenerations}{80}
\newcommand{\HypWinnerGenMin}{14}
\newcommand{\HypWinnerGenMax}{76}
\newcommand{\HypHoursMin}{1.1}
\newcommand{\HypHoursMax}{2.6}
\newcommand{\HypCpusPerActor}{2}
\newcommand{\HypCalibrations}{120}
\newcommand{\HypDriftMedianCpuTwo}{0.9 \%}
\newcommand{\CostCounters}{seven}
\newcommand{\CostWeightEdgesMinsum}{1.00}
\newcommand{\CostWeightEdgesProdsum}{1.91}
\newcommand{\CostWeightGfTwoWords}{0.28}
\newcommand{\CostWeightCandWords}{2.25}
\newcommand{\CostWeightScanElems}{0.30}
\newcommand{\CostWeightOpCalls}{0.00}
\newcommand{\CostWeightClastTicks}{0.00}
\newcommand{\CostMinsumOverProdsum}{0.52}
\newcommand{\CostFitPNinetyFive}{11.8\%}
\newcommand{\CostFitHoldoutPNinetyFive}{21.2\%}
\newcommand{\CostFitRSq}{0.998}
\newcommand{\CostFitPrograms}{252}
\newcommand{\CostSelectionMargin}{1.4}
\newcommand{\CostMarginFloor}{1.31}
\newcommand{\CostRef}{12\,924.58}
\newcommand{\CostRankCFourTau}{0.973}
\newcommand{\CostRankCFourSpearman}{0.998}
\newcommand{\CostRankMsukTau}{0.935}
\newcommand{\CostRankMsukSpearman}{0.992}
\newcommand{\CertPOneOracle}{ldpc BpDecoder; the damping referee}
\newcommand{\CertPOneResultPaper}{serial: equal on every shot (1\,024 x 6 points); flooding: 955-1\,020 / 1\,024 decisions equal, failures inside McNemar; float64 referee = ldpc on 4\,096 / 4\,096}
\newcommand{\CertPTwoOracle}{ldpc BpOsdDecoder}
\newcommand{\CertPTwoResultPaper}{corrections 1\,024 / 1\,024 and 1\,023 / 1\,024; failures equal}
\newcommand{\CertPTwoDOracle}{the damping referee; the undamped kernel via P1}
\newcommand{\CertPTwoDResultPaper}{decisions equal the referee's on every shot at 10 iterations}
\newcommand{\CertPThreeOracle}{ldpc BpOsdDecoder(osd\_cs)}
\newcommand{\CertPThreeResultPaper}{corrections 1\,024 / 1\,024 at all four points; failures equal}
\newcommand{\CertPThreeOursOracle}{none}
\newcommand{\CertPFourOracle}{none}
\newcommand{\CertPFiveOracle}{relay-bp 0.2.2, given the crate's own strengths}
\newcommand{\CertPFiveResultPaper}{kernel per shot to ten iterations; leg machinery identical on every shot any stage converges (4 x 1\,024); full decoder agrees with relay-bp's float32 as often as its float64 does, failures inside McNemar}
\newcommand{\CertPSixOracle}{none}
\newcommand{\CertPSevenOracle}{ldpc BpLsdDecoder(LSD\_0)}
\newcommand{\CertPSevenResultPaper}{predictions 1\,024 / 1\,024 at both p behind the serial BP; no invalid correction}
\newcommand{\CertPSevenDOracle}{the damping referee; P7's LSD}
\newcommand{\CertPEightOracle}{ldpc LSD\_CS 1 and 10 (nearest)}
\newcommand{\CertPEightResultPaper}{failures 50 / 49 and 0 / 0 (McNemar p = 1.0)}
\newcommand{\CertPEightDOracle}{as P7-d}
\newcommand{\CertPNineOracle}{none}
\newcommand{\CertPTenResultPaper}{predictions equal on 16\,384 / 16\,384 at four settings and both p; the Ambiguity Clustering paper's Table 1 point at 0.787x (0.768-0.807) on 524\,288 fresh shots, within 25 \% at 95 \%}
\newcommand{\CertPTenOursOracle}{none}
\newcommand{\CertPTwelveOracle}{none}

\newcolumntype{L}[1]{>{\raggedright\arraybackslash}p{#1}}
\newcommand{\ldpc}{\texttt{ldpc}}

\title{Evolving Decoding Algorithms\\for Quantum Error Correction}

\author{%
  Eduardo Dueñez$^{1,2,\dagger}$
  \quad Edgar Said Hernandez Sanchez$^{1,\dagger,\ddagger}$ \\[2pt]
  Edgar A. Duéñez-Guzmán$^{1}$
  \quad Suzanne Sadedin$^{1}$ \\[2pt]
  $^{1}$Gibran.AI
  \quad $^{2}$The University of Texas at San Antonio \\[2pt]
  $^{\dagger}$Equal contribution.
  \quad $^{\ddagger}$Corresponding author. \\[2pt]
  \texttt{\{eduardo, said, duenez, suzanne\}@gibran.ai}
}

\begin{document}

\maketitle

\begin{abstract}
Quantum computation entails robustness engineering:
 a noisy quantum state is measured continuously, and a classical algorithm (the \emph{decoder}) must infer corrections that reliably remove the noise, fast enough to keep up.
We write several published belief-propagation (BP) decoders for quantum low-density parity-check (qLDPC) codes in a \emph{Domain-Specific Language (DSL)}, and search over it with grammatical evolution.
Written in the language, the published decoders show their parameters in plain view, allowing us to re-parameterise BP+OSD (ordered statistics decoding) and BP+LSD (localized statistics decoding) decoders with damping, before any search, making each \OctDampOsdCsFrontSpeedup--\OctDampOsdZeroSpeedup$\times$ faster at a higher error rate, and BP+OSD-0 about $\SepLdpcOsdZeroTimeXPTwoD\times$ faster than the \ldpc{} package's version tuned on the $[[72,12,6]]$ bivariate-bicycle code.
Evolution over the language finds decoders that improve the time–error front.
One makes \SepCThreeErrCutVsPTwoD{} fewer logical errors than BP+OSD-0 with damping at \SepCThreeTimeXPTwoD$\times$ its time, and decoders evolved from random code are more accurate than \ldpc{}'s tuned BP+OSD-0 at $1/\SepCTenLdpcOverThis$ of its time or less.
Frozen at one operating point, the two fastest evolved decoders generalize well across noise levels, code sizes and round counts, staying at the time-error front.
Ultimately, we demonstrate that capturing expert knowledge within a DSL creates a powerful intermediate layer—one that effectively bridges human intuition and machine-driven optimization to unlock highly efficient quantum error correction.
\end{abstract}

\section{Introduction}
\label{sec:intro}

A quantum computer is a machine whose very state is in constant noisy decay, and whose scalability rests critically on recovery mechanisms of \emph{quantum error correction (QEC)} (for an accessible introduction, see \citealp{roffe2019quantum}). In QEC,
measurements detect which errors have occurred; a classical algorithmic \emph{decoder} infers which corrections to apply.
The better the decoder, the more robust the system is.

A quantum state encodes data that cannot be inspected directly: observation is a destructive process;
monitoring is therefore indirect, relying on repeated measurements of parity checks on ancilla qubits.
Noise causes these checks to trigger, and the resulting pattern is called a \emph{syndrome}.
The decoder's job, given a stream of syndromes, is to infer appropriate corrections in real time, round after round.
This presents two distinct engineering challenges.
First, computational intractability~\citep{iyer2015hardness}:
every practical decoder must be a heuristic balancing accuracy with real-time performance.
Second, degeneracy: distinct errors may produce the same syndrome and require equally valid corrections. This symmetry stalls message-passing inference~\citep{poulin2008iterative}, requiring the decoder to algorithmically break such ties.

To overcome the limitations of human-designed heuristics, we carry out an evolutionary search for decoder algorithms.
The substrate of evolution is a \emph{Domain-Specific Language (DSL)}:
a small vocabulary of typed operations in which the published decoders of one family—belief propagation (BP, recognizable to the ML community as a graph-based message-passing algorithm for probabilistic graphical models) plus a tie-breaking post-processor—are short programs (§\ref{sec:setup}, App.~\ref{app:dsl}).
The DSL plays two roles: it defines the space of algorithms explored by the search, and simultaneously becomes the lexicon recording all such algorithms found: it is a common layer that human experts can understand, and computers can explore mechanically.
Regardless of the manner in which the code is produced or found—by evolution, by human researchers, or by language models—it is a tangible artefact useful to two audiences:
humans can read and compare it against other algorithms; computers can parse and execute it, measuring its effectiveness.
The DSL is thus not a convenience but the decisive expert-knowledge input: domain knowledge enters through the representation; the search itself stays generic.
Prior evolutionary work tunes weights inside a fixed Belief Propagation with Ordered Statistics Decoding decoder (BP+OSD)~\citep{kwak2025evolutionary}; ours writes the algorithm.

The paper makes two contributions.
\emph{Contribution~1 (methodological)}: writing the state of the art in a language exposes what its implementations hide.
By expressing BP+OSD and BP with Localized Statistics Decoding (BP+LSD) as compositions of named and typed steps, their parameters are brought into plain view. Simply re-parameterising them with damping made them significantly faster, at a higher error rate, even before any search is applied (§\ref{sec:c1}).
\emph{Contribution~2 (algorithmic)}: We discover evolved decoders that break the Pareto front of speed versus accuracy —running faster than the reference decoder while producing fewer errors— including some evolved entirely from random code (§\ref{sec:c2}).
Every evolved decoder reported here was produced by grammatical evolution alone. There is no language model in the variation loop.
These results provide a counterexample to the view that genetic programming~\citep{koza1992genetic} is ineffective on real algorithm-discovery problems~\citep{oneill2010open}, as well as to the prevailing narrative in recent LLM-driven program discovery~\citep{romeraparedes2024funsearch,novikov2025alphaevolve} that classical variation is the weak link.

\section{Experiment setup}
\label{sec:setup}

We focus on the $[[72,12,6]]$ bivariate-bicycle code~\citep{bravyi2024high} under circuit-level noise
(physical error rate $p = 0.004$, six rounds, $Z$ basis)—a configuration we designate as the exam-point—with the error model extracted by the Python library Stim~\citep{gidney2021stim}.
A syndrome acts as the input feature vector, and one \emph{shot} represents a single inference pass. A shot is decoded correctly when all twelve encoded bits are recovered, making the logical error rate (LER)—the fraction of failed shots—our primary classification error metric.

We designed the DSL by reading the published BP+ family of decoders (the panel) and factoring them into reusable operations, each an algorithmic step used by at least one of the published decoders: \HypDslOps{} typed operations in eight groups, with enumerated tags for rules, schedules and orderings, and a constant prior drawn from the published operating points (Appendix~\ref{app:dsl}).
An evolved decoder is an ordinary typed function \texttt{solve(state)} $\to$ \texttt{state}, called once per shot; its control flow is written by evolution itself.
Panel and search space are one family by construction, so neural, matching-based and tensor-network decoders lie outside both (Appendix~\ref{app:panel}).

{\footnotesize
\begin{verbatim}
def solve_bp_osd_cs(state):
    state = run_bp(state, BP_ITERS, BpRule.MIN_SUM, BP_ALPHA, 0.0,
                   Schedule.PARALLEL, BP_GAMMA)           
    if bp_converged(state):
        return state
        
    ranking = rank_faults(state, RankKey.POSTERIOR, Order.ASCENDING)
    state = most_reliable_basis(state, ranking)
    state = solve_on_basis(state)
    state = keep_candidate(state)
    state = enumerate_patterns(state, OSD_CS_WEIGHT, OSD_CS_TOP, Scope.GLOBAL)
    state = select_best(state, SelectKey.PRIOR_WEIGHT)
    return state
\end{verbatim}
}
\noindent{\small\textbf{Listing 1: BP+OSD-CS as written in the DSL (our damped variant).}}

Published comparators use the authors' own packages at their tuned settings where available: \ldpc{}'s BP+OSD-0 and BP+OSD-CS~\citep{roffe2020decoding,panteleev2021degenerate,fossorier1995soft}, and Relay-BP~\citep{muller2025improved}. For Ambiguity Clustering (BP+AC)~\citep{wolanski2024ambiguity}, which lacks an open-source implementation, we use our DSL implementation of its published stages, certified against an independent re-implementation (Appendix~\ref{app:panel}).

Times in the body are clean wall-clock milliseconds per shot on a benchmark workstation (Apple M3 Ultra, single process, idle), with ratios against the reference decoder of §\ref{sec:c1} unless another is named.

The engine is MAP-Elites~\citep{mouret2015illuminating,pugh2016quality} utilizing a 1-dimensional behavioral characterization (computational cost) with one cell per cost bin, which forces the quality-diversity engine to act as an explicit Pareto-front discoverer. Parents are drawn by tournaments of \HypTournament{} over the cells, each generation \HypPrograms{} programs scored on a \HypTrainShots-shot training draw. To efficiently navigate the fitness landscape, we apply an importance sampling technique: the draw over-samples the hard instances where plain BP fails and reweights each by its inverse sampling probability. This focuses the selection gradient on informative edge cases while keeping the objective function unbiased.
Selection ranks on accuracy plus a weighted cost term limited by per-shot budgets (Appendix~\ref{app:protocol}).

The campaign ran \HypRuns{} runs: two founder arms (\emph{seeded} with \HypSeedMembers{} published compositions, or \emph{random code}) times two cost axes (the work proxy, or wall-clock calibrated per worker) times \HypReplicates{} replicates, each run for \HypGenerations{} generations (\HypHoursMin–\HypHoursMax{} h on \HypWorkers{} actors); the winner of each run first appeared between generations \HypWinnerGenMin{} and \HypWinnerGenMax{}.
The evolved decoders are named C01 to C15: the champion each run selected on the validation split (C01 to C12; C01 and C02 duplicate C03 and C09 and are not reported separately), and three further validation finalists from the archive (C13 to C15; App.~\ref{app:protocol}).

Three dataset splits were generated: \emph{train} (drives selection), \emph{validation} (screening and candidate naming) and \emph{final}.

The final split is uniform, unweighted, and strictly read once, exclusively for subjects named in advance, against dominance bars pre-registered before the campaign ran (Appendix~\ref{app:protocol}). Comparisons are paired on common shots and adjudicated by McNemar tests~\citep{mcnemar1947note,dietterich1998approximate}.

In the literature, qLDPC decoders are already compared on accuracy and time together, matched to BP+OSD's accuracy and then timed~\citep{wolanski2024ambiguity,hillmann2025localized} or faster and more accurate at once~\citep{ye2026beam}. Recent benchmarking names the criterion: a decoder is on the Pareto frontier when none is both faster and more accurate~\citep{manor2026simtoreal}. We adopt it as the standard for a \emph{discovered} recovery mechanism: \emph{Pareto competitiveness} against the published panel and not record-setting accuracy. Improving robustness (accuracy) competes for the compute budget (cost) of the recovery loop, which in a fault-tolerant stack is a hard real-time constraint. Pressure on the trade-off is what keeps discovered mechanisms deployable rather than merely accurate.

\section{Results}
\label{sec:results}

\paragraph{Contribution 1: writing the state of the art in a language exposes damping.} \label{sec:c1} Writing the BP-plus-post-processor family in the DSL puts every parameter of every published decoder in one place, and one of them, the damping of the BP front end~\citep{muller2025improved}, is a knob the \ldpc{} package does not expose and cannot be tuned.

In loopy graphical models, standard BP often oscillates and fails to converge; damping acts as a momentum term that stabilizes message passing. The ability to modify this parameter makes the computationally inexpensive—but normally unstable—flooding min-sum schedule viable. Tuning it results in a decoder comparable to \ldpc{}'s BP+OSD-0; while it incurs a higher LER (\SepLdpcOsdZeroLer{} to \SepPTwoDLer{}), it runs in roughly $1/\SepLdpcOsdZeroTimeXPTwoD$ of the time, effectively leveraging the speed-for-accuracy tradeoff inherent to this representation.
We call this tuned decoder P2-d and make it the cost reference ($1.0\times$) for everything that follows, chosen because it is the fastest decoder in its family.
Table~\ref{tab:damping} isolates the parameter effect across the decoders in \ldpc{} by running both versions in the DSL's engine.
Every damped member is faster, by \OctDampOsdCsFrontSpeedup$\times$ to \OctDampOsdZeroSpeedup$\times$, and every damped member is less accurate.

\begin{table}[t]
  \centering
  \caption{%
    Contribution~1: each decoder published in the \ldpc{} library, at its published settings and damped, both run in the DSL's engine.
    Time is clean ms per shot, one single-threaded process; LER is on the final split; $z$ is McNemar, positive when the published member is right more often.
    Settings and the timing protocol are in Appendix~\ref{app:damping}.
  }
  \label{tab:damping}
  \small
  \setlength{\tabcolsep}{4pt}
  \begin{tabular}{llrrrll}
    \toprule
    family & \multicolumn{2}{c}{ms per shot} & speed-up & LER, pub.\ $\to$ damped & $z$ \\
    \cmidrule(lr){2-3}
    & pub. & damped & & & \\
    \midrule
    BP+OSD-0 & \OctDampPTwoMs & \OctDampPTwoDMs & \textbf{\OctDampOsdZeroSpeedup$\times$} & \OctDampPTwoLer{} $\to$ \OctDampPTwoDLer & \OctDampOsdZeroZ \\
    BP+LSD-0 & \OctDampPSevenMs & \OctDampPSevenDMs & \textbf{\OctDampLsdZeroSpeedup$\times$} & \OctDampPSevenLer{} $\to$ \OctDampPSevenDLer & \OctDampLsdZeroZ \\
    BP+LSD-$\mu$ & \OctDampPEightMs & \OctDampPEightDMs & \textbf{\OctDampLsdMuSpeedup$\times$} & \OctDampPEightLer{} $\to$ \OctDampPEightDLer & \OctDampLsdMuZ \\
    BP+OSD-CS & \OctDampPThreeMs & \OctDampPThreeDMs & \textbf{\OctDampOsdCsFrontSpeedup$\times$} & \OctDampPThreeLer{} $\to$ \OctDampPThreeDLer & \OctDampOsdCsFrontZ \\
    \bottomrule
  \end{tabular}
\end{table}

\paragraph{Contribution 2: evolved decoders that break the Pareto front.} \label{sec:c2} Sixteen finalists were drawn from the four experimental arms. Four of these (\OctAdjStretchWinners) met the stretch bar: not Pareto-dominated by any panel member, more accurate than the reference, and cheaper than it by at least the \OctAdjMargin$\times$ margin.

\begin{figure}[t]
  \centering
  \includegraphics[width=\textwidth]{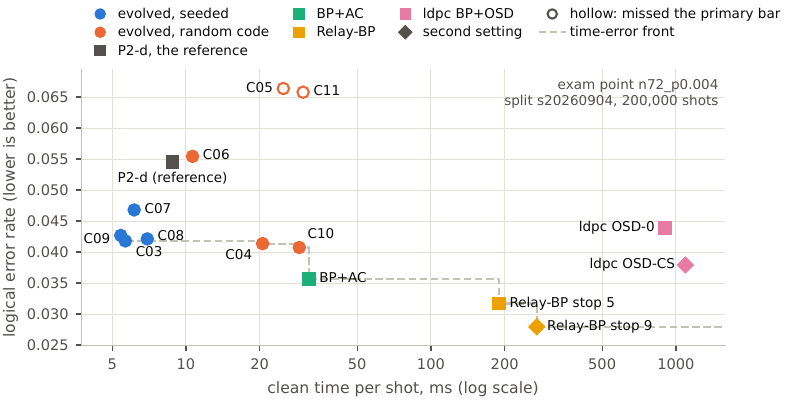}
  \caption{%
    Logical error rate against clean time per shot (log scale) at the benchmark's exam-point ($n = 72$, $p = 0.004$, six rounds; \SepShots{} shots); down and left is better.
    C12 (LER \SepCTwelveLer) lies outside the frame; hollow marks missed the primary bar.
    The dashed line is the time–error front computed from the drawn rows: C09, C03, C04, C10, BP+AC, Relay-BP at stop 5 and at stop 9.
  }
  \label{fig:pareto}
\end{figure}

Of these, C03 makes \SepCThreeErrCutVsPTwoD{} fewer errors than the reference at \SepCThreeTimeXPTwoD$\times$ its time ($z = \SepCThreeZVsPTwoD$, at the exam-point of Figure~\ref{fig:pareto}).
Measured against \ldpc{}'s tuned BP+OSD-0 instead, the same decoder runs at $1/\SepCThreeLdpcOverThis$ of its time with \SepCThreeErrCutVsLdpcOsdZero{} fewer errors ($z = \SepCThreeZVsLdpcOsdZero$).

Figure~\ref{fig:pareto} draws every benchmarked row; the front computed from them runs C09, C03, C04, C10, BP+AC, Relay-BP at stop 5 and at stop 9, so the four fastest decoders on it are evolved.
Ambiguity Clustering and Relay-BP reach lower error rates, at about \SepAcTimeXCThreeWord{} and \SepRelayFiveTimeXCThreeWord{} times C03's time ($z = \SepRelayFiveZVsAc$ for Relay-BP over BP+AC).
Relay-BP at its tuned point is more accurate than every evolved decoder in this paper —at stop 9 it errs on \SepRelayNineLer{} of shots at \SepRelayNineMs{} ms.

The seeded winners are \NovWordingCRows{}: C03 and C09 run BP+AC's own stages with min-sum BP in place of product-sum (Appendix~\ref{app:code}), C07 and C08 add a degeneracy cut (a published op) before BP+AC's search, and at the accurate end C14 retunes BP+AC's constants to beat published BP+AC on the final split (\OctAdjETwentySevenLer{} vs.\ \OctAdjPTenLer{}, $z = \OctAdjETwentySevenZVsPTen$) at \OctAdjETwentySevenTimeXAc$\times$ its time (Listing~3; Table~\ref{tab:slate}).

From random code, evolution re-found BP+OSD-0 and tuned its BP stage (damped parallel product-sum), reaching the front past \ldpc{}'s tuned BP+OSD-0: C10 makes \SepCTenErrCutVsPTwoD{} fewer errors than the reference ($z = \SepCTenZVsPTwoD$) and is more accurate than \ldpc{}'s tuned BP+OSD-0 ($z = \SepCTenZVsLdpcOsdZero$) at $1/\SepCTenLdpcOverThis$ of its time, and C04 is more accurate ($z = \SepCFourZVsLdpcOsdZero$) at $1/\SepCFourLdpcOverThis$.
Both are Pareto-undominated past \ldpc{}'s BP+OSD-0 and never past BP+AC, which at about C10's time has the lower error rate ($z = \SepCTenZVsAc$ for C10).

\begin{figure}[t]
  \centering
  \includegraphics[width=\textwidth]{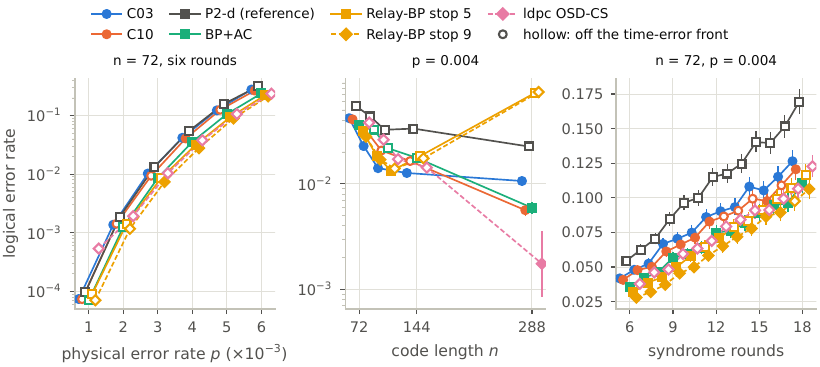}
  \caption{%
    Logical error rate with every decoder frozen at its exam-point settings, against physical error rate at $n = 72$ (left), code length at $p = 0.004$ (centre) and syndrome rounds at $n = 72$, $p = 0.004$ (right), with Wilson 95\% intervals.
    A filled mark is on the time–error front at that point over every benchmarked row, and a hollow one is off it; C03 is filled at every point.
    Shot counts run from \GridShotsMax{} ($p = 0.001$) down to \GridPSixShots{} ($p = 0.006$) and \GridREighteenShots{} (18 rounds); at $n = 288$, \ldpc{}'s BP+OSD-CS was decoded on \GridNTwoHundredEightyEightLdpcOsdCsShots{} of the point's \GridNTwoHundredEightyEightShots{} shots.
    The error-rate axis is logarithmic on the left and centre panels and linear on the right.
  }
  \label{fig:generalisation}
\end{figure}

Frozen at their exam-point, C03 and C09 are on the time–error front at all \GridPoints{} benchmark points, with C03 strictly dominating the reference—making fewer errors in less time—at every point and beating published BP+AC at $n = 90, 108, 144$ (Figure~\ref{fig:generalisation}).
Over the benchmarked rows, BP+AC is on the front at \GridAcFront{} points, Relay-BP at \GridRelayNineFront{} (stop 9) and \GridRelayFiveFront{} (stop 5), C04 at \GridCFourFront{} and C10 at \GridCTenFront{}, and the reference at none.
At $n = 144$, C03 makes \GridNOneHundredFortyFourCThreeErrors{} errors in \GridNOneHundredFortyFourShots{} shots at \GridNOneHundredFortyFourCThreeMs{} ms, against BP+AC's \GridNOneHundredFortyFourAcErrors{} at \GridNOneHundredFortyFourAcMs{} ms, Relay-BP's \GridNOneHundredFortyFourRelayFiveErrors{} at \GridNOneHundredFortyFourRelayFiveMs{} ms (stop 5) and the reference's \GridNOneHundredFortyFourPTwoDErrors{} at \GridNOneHundredFortyFourPTwoDMs{} ms.
The fronts compare point estimates, and shot counts vary by point (Appendix~\ref{app:transfer}), so the counts are not significance statements.

\section{Conclusion}
\label{sec:conclusion}

This work shows that evolutionary search can discover efficient recovery mechanisms for quantum error correction, provided the domain expertise is encoded into a Domain-Specific Language (DSL).

The methodological contribution came before any search: writing the published decoders in a language put their parameters in plain view, and re-parameterising BP+OSD with damping ran it \OctDampOsdZeroSpeedup$\times$ faster within the same engine—and at about a hundredth of the packaged time—at a higher error rate, a trade exposed by the representation.

The algorithmic contribution is driven by the search itself. Operating from both seeded and purely random code, grammatical evolution discovered decoders that are faster than the reference while producing fewer errors. Crucially, the fastest of these evolved decoders hold the fast end of the time-error front and remain there—even when frozen—across a \GridPoints-point benchmark spanning various noise levels, code sizes, and rounds. Ambiguity Clustering and Relay-BP remain more accurate, at several times their cost.

Ultimately, the transferable lesson of this work lies in the representation, not in the optimizer. A Domain-Specific Language serves as a common layer whose programs are natively readable by people, executable by machines, and discoverable by evolution. We expect this framework to outlast our specific search, directly challenging the assumption that classical variation is obsolete. The true bottleneck in automated algorithmic discovery is not a lack of sophisticated generative models, but the absence of tangible, auditable environments; when domain knowledge is properly embedded into the substrate, evolution alone is enough to move the fast end of the state of the art.

\label{sec:endbody}

\bibliographystyle{plainnat}
\bibliography{refs}

\appendix

\section{The DSL}
\label{app:dsl}

The language is thirty-eight typed operations in eight groups, plus enumerated tags (BP rule, message schedule, orderings, ranking keys, scopes, sectors) and a constant prior.
It was designed by reading the published BP+ family algorithms and factoring them into reusable steps; each operation (named and typed) is an algorithmic step used by at least one of the reference decoders in the published literature.

Table~\ref{tab:dsl} lists the DSL vocabulary.
Every operation is executed natively by a Rust engine on which an evolved program runs unchanged, and each benchmarked champion was also ported to a standalone native decoder, certified against the evolved program on the \SepShots{} shots of the exam-point with zero differing predictions.

\begin{table}[ht]
  \centering
  \caption{The 38 operations, by group.  Every published panel member is expressible as a short composition; Appendix~\ref{app:code} prints three decoders in this vocabulary.}
  \label{tab:dsl}
  \small
  \begin{tabular}{llL{8.2cm}}
    \toprule
    group & n & operations \\
    \midrule
    A prepare & 9 & \texttt{init\_llrs}, \texttt{scale\_llrs}, \texttt{sparsify\_dem}, \texttt{permute\_problem}, \texttt{mask\_columns}, \texttt{boost\_priors\_from\_syndrome}, \texttt{use\_check\_basis}, \texttt{inactivate\_check}, \texttt{cut\_degenerate\_supports} \\
    B engine & 5 & \texttt{run\_bp}, \texttt{bp\_converged}, \texttt{reseed\_priors\_from\_posteriors}, \texttt{rank\_by\_oscillation}, \texttt{valid\_candidates} \\
    C ranking & 5 & \texttt{rank\_faults}, \texttt{most\_reliable\_basis}, \texttt{decimate}, \texttt{decimate\_and\_continue}, \texttt{rank\_checks} \\
    D clusters & 7 & \texttt{seed\_clusters\_from\_syndrome}, \texttt{grow\_clusters}, \texttt{syndrome\_guided\_reduce}, \texttt{reduce\_marked\_rows}, \texttt{enrich\_blocks}, \texttt{clusters\_valid}, \texttt{build\_forest} \\
    E solvers & 3 & \texttt{solve\_on\_basis}, \texttt{incremental\_eliminate}, \texttt{forced\_logical} \\
    F local search & 2 & \texttt{enumerate\_patterns}, \texttt{vote\_logical\_classes} \\
    G control & 5 & \texttt{keep\_candidate}, \texttt{select\_best}, \texttt{branch\_frontier}, \texttt{advance\_frontier}, \texttt{prune\_frontier} \\
    H assembly & 2 & \texttt{assemble\_cluster\_solutions}, \texttt{map\_back} \\
    \bottomrule
  \end{tabular}
\end{table}

Three design rules at the core:
first, typed enumerations instead of numbers wherever the literature has a discrete choice (rule, schedule, ordering), so that mutation moves among meaningful alternatives;
second, the constant prior is the literature's own operating points —BP iteration caps, damping factors, sweep depths and masking fractions are the values the papers use— and every tuned panel parameter is reachable as a listed constant, a fairness rule enforced by test;
third, exclusions are stated, not silent: one catalogue operation (a per-shot linear-programming engine) is deliberately absent on per-shot cost grounds.
The DSL vocabulary spans the family, and variation respects the types:
mutation and crossover only ever substitute type-compatible subtrees and constants, so every offspring parses and runs by construction, with no syntactic-repair step.

\section{Two evolved decoders in full, with Ambiguity Clustering for contrast}
\label{app:code}

The audit that reads an evolved program has three clauses, scored by an automated audit against every panel member: \emph{structural} (the canonical call multiset differs), \emph{behavioural} (shot-for-shot predictions differ after ablation) and \emph{discordance} (at least one decision differs).

A simple ablation removes whole top-level statements only, so a live statement carrying dead wrappers keeps them and the structural clause counts each; we therefore also prune below the statement —delete a statement at any depth, unwrap a call to its argument, flatten a constant branch— keeping an edit only when every prediction stays identical to the evolved program's, and confirm each pruned form on \NovConfirmShots{} validation shots.
Table~\ref{tab:novelty} gives the verdict for the \NovSubjectsListed{} decoders this paper names.

\begin{table}[ht]
  \centering
  \caption{%
    What the evolved decoders are, after pruning: calls in the program as evolved and after pruning, the nearest panel member and the call-multiset distance to it after pruning, and the verdict.
    Every pruned form is prediction-identical to its evolved program on \NovConfirmShots{} validation shots; ``unseeded'' is the random-code arm.
  }
  \label{tab:novelty}
  \footnotesize
  \setlength{\tabcolsep}{3pt}
  \begin{tabular}{llrrlrL{4.2cm}}
    \toprule
    decoder & arm & \multicolumn{2}{c}{calls} & nearest & dist. & verdict \\
    \cmidrule(lr){3-4}
    & & evolved & pruned & & & \\
    \midrule
    C03 & \NovCThreeArm & \NovCThreeCallsVerbatimN & \NovCThreeCallsPrunedN & \NovCThreeNearest & \NovCThreeDistPruned & \NovCThreeVerdictFull \\
    C09 & \NovCNineArm & \NovCNineCallsVerbatimN & \NovCNineCallsPrunedN & \NovCNineNearest & \NovCNineDistPruned & \NovCNineVerdictFull \\
    C07 & \NovCSevenArm & \NovCSevenCallsVerbatimN & \NovCSevenCallsPrunedN & \NovCSevenNearest & \NovCSevenDistPruned & \NovCSevenVerdictFull \\
    C08 & \NovCEightArm & \NovCEightCallsVerbatimN & \NovCEightCallsPrunedN & \NovCEightNearest & \NovCEightDistPruned & \NovCEightVerdictFull \\
    C14 & \NovETwentySevenArm & \NovETwentySevenCallsVerbatimN & \NovETwentySevenCallsPrunedN & \NovETwentySevenNearest & \NovETwentySevenDistPruned & \NovETwentySevenVerdictFull \\
    C04 & \NovCFourArm & \NovCFourCallsVerbatimN & \NovCFourCallsPrunedN & \NovCFourNearest & \NovCFourDistPruned & \NovCFourVerdictFull \\
    C10 & \NovCTenArm & \NovCTenCallsVerbatimN & \NovCTenCallsPrunedN & \NovCTenNearest & \NovCTenDistPruned & \NovCTenVerdictFull \\
    \bottomrule
  \end{tabular}
\end{table}

{\footnotesize
\begin{verbatim}
def solve(state):
    state = run_bp(state, 9, BpRule.MIN_SUM, 0.7710300216438394,
                   0.003077120912963548, Schedule.PARALLEL,
                   0.3106797689580642)
    state = reduce_marked_rows(state, True)
    state = enrich_blocks(state, 0.04118582037119639, True)
    state = enumerate_patterns(state, 1, 4000, Scope.BLOCK_VOTE)
    return state
\end{verbatim}
}

\noindent{\small\textbf{Listing 2: C03, the fast end, pruned.}
Ambiguity Clustering's own stages —BP, elimination on the marked rows, enrichment of each cluster's block, a per-cluster search— with min-sum BP in place of product-sum, a narrower enrichment and a weight-1 search, in \NovCThreeCallsPruned{} calls.
BP+AC's early exit after BP is gone: it changes no decision of C03's on \NovConfirmShots{} shots.
Every operation is one of the thirty-eight in Appendix~\ref{app:dsl}, and the whole function is called once per shot.

Table~\ref{tab:ablation} says where the speed comes from. Putting product-sum BP back into either decoder roughly doubles its time and buys back part of the accuracy; restoring BP+AC’s enrichment width or its search weight costs time and does not help, and BP+AC’s early exit after BP, which C09 dropped, changes no decision. Substituting min-sum for product-sum BP inside BP+AC's cluster post-processor is the mechanism that unlocks the fast end of the front.}

{\footnotesize
\begin{verbatim}
def solve(state):
    state = run_bp(state, 31, BpRule.PRODUCT_SUM, 0.625, 0.0,
                   Schedule.PARALLEL, 0.4123303093240179)
    state = reduce_marked_rows(state, True)
    state = enrich_blocks(state, 0.03269060068025888, True)
    state = enumerate_patterns(state, 2, 4000, Scope.BLOCK_VOTE)
    return state

def solve_ambiguity_clustering(state):   # BP+AC as published, for contrast
    state = run_bp(state, 45, BpRule.PRODUCT_SUM, 0.625, 0.0,
                   Schedule.PARALLEL, 0.3)
    if bp_converged(state):
        return state
    state = reduce_marked_rows(state, True)
    state = enrich_blocks(state, 0.1, True)
    state = enumerate_patterns(state, 2, 4000, Scope.BLOCK_VOTE)
    return state
\end{verbatim}
}

\noindent{\small\textbf{Listing 3: C14, pruned, above Ambiguity Clustering as published.}
The same four stages with three constants moved: \NovETwentySevenSweeps{} parallel product-sum sweeps against BP+AC's \NovAcSweeps{}, damping $\gamma = \NovETwentySevenGamma$ against \NovAcGamma{} and enrichment $\kappa = \NovETwentySevenKappa$ against \NovAcKappa{}; BP+AC's early exit is inert for C14.
Setting the three back one at a time on \NovConfirmShots{} validation shots, the damping carries \NovETwentySevenErrorsFromGamma{} of the \NovETwentySevenErrorsBelowAc{} errors by which C14 beats BP+AC, and the sweep cut carries the saving in cost (\NovETwentySevenCostXAc$\times$ BP+AC's in the work proxy).
Even without altering BP+AC's four-stage structure, evolutionary tuning of these three parameters yields a decoder that strictly dominates the published configuration in both speed and accuracy.}

\begin{table}[ht]
  \centering
  \caption{%
    Where the speed comes from: C03 and C09 with one of BP+AC's settings put back at a time, on \SepAblShots{} shots of split \SepSplit{}, timed on the benchmark workstation.
    $\times$ is time over the evolved decoder's and $z$ is McNemar against it (positive: the variant is more accurate).
    The last two rows are BP+AC as published and the reference on the same shots, against C03.
  }
  \label{tab:ablation}
  \footnotesize
  \setlength{\tabcolsep}{3pt}
  \begin{tabular}{lrrrrrrrr}
    \toprule
    & \multicolumn{4}{c}{C03} & \multicolumn{4}{c}{C09} \\
    \cmidrule(lr){2-5} \cmidrule(lr){6-9}
    change put back & LER & ms & $\times$ & $z$ & LER & ms & $\times$ & $z$ \\
    \midrule
    none, as evolved & \SepAblCThreeLer & \SepAblCThreeMs & 1.0 & & \SepAblCNineLer & \SepAblCNineMs & 1.0 & \\
    product-sum BP & \SepAblCThreeProductSumLer & \SepAblCThreeProductSumMs & \SepAblCThreeProductSumTimeXBase & \SepAblCThreeProductSumZ & \SepAblCNineProductSumLer & \SepAblCNineProductSumMs & \SepAblCNineProductSumTimeXBase & \SepAblCNineProductSumZ \\
    enrichment $\kappa = 0.1$ & \SepAblCThreeKappaLer & \SepAblCThreeKappaMs & \SepAblCThreeKappaTimeXBase & \SepAblCThreeKappaZ & \SepAblCNineKappaLer & \SepAblCNineKappaMs & \SepAblCNineKappaTimeXBase & \SepAblCNineKappaZ \\
    search weight 2 & \SepAblCThreeWeightTwoLer & \SepAblCThreeWeightTwoMs & \SepAblCThreeWeightTwoTimeXBase & \SepAblCThreeWeightTwoZ & --- & & & \\
    early exit after BP & --- & & & & \SepAblCNineEarlyExitLer & \SepAblCNineEarlyExitMs & \SepAblCNineEarlyExitTimeXBase & \SepAblCNineEarlyExitZ \\
    all three at once & \SepAblCThreeAllThreeLer & \SepAblCThreeAllThreeMs & \SepAblCThreeAllThreeTimeXBase & \SepAblCThreeAllThreeZ & --- & & & \\
    \midrule
    BP+AC as published & \SepAblAcLer & \SepAblAcMs & \SepAblAcTimeXBase & \SepAblAcZ & & & & \\
    P2-d, the reference & \SepAblPTwoDLer & \SepAblPTwoDMs & \SepAblPTwoDTimeXBase & \SepAblPTwoDZ & & & & \\
    \bottomrule
  \end{tabular}
\end{table}

\section{The panel}
\label{app:panel}

Table~\ref{tab:panel} names the panel's rows and how each was certified.
The published members are certified per shot against \ldpc{}~\citep{roffe2020decoding} (BP+OSD-0, BP+OSD-CS, BP+LSD) and against the Relay-BP package (Relay-BP, at the authors' tuned point); Ambiguity Clustering, which has no open-source implementation, is certified against an independent re-implementation of its published stages and against the Ambiguity Clustering paper's own reported point.
Rows without an oracle are our variants or regime rows, and are labelled as such.
The damped front ends have no package oracle and are certified against a float64 damping referee.
P13 is structurally void at this operating point, and the primary bar counts dominance over the other members only.

\begin{table}[ht]
  \centering
  \caption{%
    The panel and its certification.
    ``Oracle'' is the independent implementation each row was checked against, and ``result'' what the check found on the shots stated.
    Rows marked ``---'' have no oracle: they are our variants, regime rows, or rows not carried.
  }
  \label{tab:panel}
  \scriptsize
  \setlength{\tabcolsep}{3pt}
  \renewcommand{\arraystretch}{1.25}
  \begin{tabular}{L{1.05cm}L{2.3cm}L{2.2cm}L{2.65cm}L{4.6cm}}
    \toprule
    \multicolumn{3}{c}{member} & oracle & result \\
    \cmidrule(lr){1-3}
    id & decoder & reference & & \\
    \midrule
    P1 & BP & \citep{poulin2008iterative} & \CertPOneOracle & \CertPOneResultPaper \\
    P2 & BP+OSD-0 at \ldpc{}'s tuned point & \citep{panteleev2021degenerate} & \CertPTwoOracle & \CertPTwoResultPaper \\
    P2-d & damped (the reference) & & \CertPTwoDOracle & \CertPTwoDResultPaper \\
    P3 & BP+OSD-CS & \citep{panteleev2021degenerate,fossorier1995soft} & \CertPThreeOracle & \CertPThreeResultPaper \\
    P3-ours & our OSD-CS sweep & & \CertPThreeOursOracle & our variant \\
    P4 & BPGD, as published & \citep{yao2023belief} & \CertPFourOracle & regime row: a code-capacity method with no circuit-level claim \\
    P5 & Relay-BP, our port & \citep{muller2025improved} & \CertPFiveOracle & \CertPFiveResultPaper \\
    P6 & BP+SI-lite & \citep{crest2022stabilizer} & \CertPSixOracle & our variant \\
    P7 & BP+LSD-0 & \citep{hillmann2025localized} & \CertPSevenOracle & \CertPSevenResultPaper \\
    P7-d & damped & & \CertPSevenDOracle & damped front end as P1 \\
    P8 & BP+LSD-$\mu$ & \citep{hillmann2025localized} & \CertPEightOracle & \CertPEightResultPaper \\
    P8-d & damped & & \CertPEightDOracle & damped front end as P1 \\
    P9 & BP+OTF & \citep{iolius2026almost} & \CertPNineOracle & the published forest on the raw error model \\
    P10 & BP+AC, as published & \citep{wolanski2024ambiguity} & an independent re-implementation; the Ambiguity Clustering paper's Table~1 & \CertPTenResultPaper \\
    P10-ours & BP+AC, our variant & & \CertPTenOursOracle & our variant \\
    P11 & BP+BS & \citep{ye2026beam} & --- & not carried \\
    P12 & BP+DC & \citep{tsubouchi2025degeneracy} & \CertPTwelveOracle & one cut per generator, BP restarted once \\
    P13 & BP+CA & \citep{crest2024inherent} & --- & structurally void here \\
    \bottomrule
  \end{tabular}
\end{table}

\paragraph{What the panel does not contain.}
The DSL factors the BP-plus-post-processor family and the panel is drawn from it, so three decoder families are absent by construction rather than by selection: neural decoders such as AlphaQubit~\citep{bausch2024learning}, which learn a decoding policy rather than compose one and are reported on surface-code data; minimum-weight perfect matching~\citep{higgott2025sparse}, which is not defined on a bivariate-bicycle code's hyperedge error model without a further reduction; and tensor-network maximum-likelihood decoding~\citep{bravyi2014efficient}, whose per-shot cost sits far outside the real-time budget this campaign selects under.
They bound the claim rather than weaken it: every comparison in this paper is against the family a $[[72,12,6]]$ decoder is built from today, and no claim here is a claim about the decoder literature as a whole.
One neighbour is worth naming separately, since it is the closest methodological one: \citet{kwak2025evolutionary} fit trainable per-message BP weights —the neural-BP parameterisation of \citet{liu2019neural}, fitted without gradients— by differential evolution, and pair them with OSD.
Their decoder is inside this family —it is a BP+OSD— but its weight vector has no counterpart in our constant space, whose BP engine carries scalar normalisation, offset and memory constants only, so it is not a panel row we could implement faithfully.
The contrast is the one this paper is about: evolution over the parameters of a fixed algorithm, against evolution over the algorithm. We did not run it as a comparator.

\section{Evaluation protocol details}
\label{app:protocol}

\paragraph{Splits and the two exam-points.}
Each dataset has three splits: train (drives selection), validation (champion naming and the held-out readings) and \emph{final}.
The adjudication reads the final split \OctAdjSplit{} (payload \OctAdjDigest), which this campaign never read; Table~\ref{tab:damping} is read there too.
The benchmark's exam-point is split \SepSplit{} (payload \SepDigest), where Figures~\ref{fig:pareto} and~\ref{fig:generalisation} and the ablation are read.

\paragraph{The enriched training pool.}
At this operating point most uniformly drawn shots are decoded by plain BP alone and so carry no signal about the post-processing under selection.
Training shots are therefore drawn from two strata —the pool on which plain BP fails, and the remainder— and each draw is weighted by its inverse sampling probability, so selection concentrates on discriminating shots while every reported accuracy remains an unbiased estimate for the uniform distribution.
The final and benchmark splits are uniform and unweighted.

\paragraph{Budgets.}
Selection ranks on accuracy plus a cost term of weight $W^{*} = \HypWStar$, the weight at which P2-d and Ambiguity Clustering rank equal, chosen on a sweep over held-out validation shots before the reported runs.
Four per-shot budgets bound what one shot may spend, and one cap bounds what an evaluation may spend on average (Table~\ref{tab:budgets}).
An operation called with its budget spent does nothing, and the shot is decoded from the state the program has reached by then; an evaluation past the work cap scores as a failure.
Column operations and the cap are twice the costliest seed's measured usage on the hard shots the training pool concentrates on; the candidate budget truncates two seeds' enumeration at no measured cost to their accuracy (no discordant shot for one, McNemar $p = 1.0$ for the other); sweeps and operation calls are kept wider than that rule's 734 and 88, so that evolved multi-round BP is not truncated.

\begin{table}[ht]
  \centering
  \caption{The campaign's budget set (\HypBudgetSet). Work is in the proxy of Appendix~\ref{app:work}, relative to the reference P2-d's.}
  \label{tab:budgets}
  \small
  \begin{tabular}{lrL{7.4cm}}
    \toprule
    budget & limit & bounds \\
    \midrule
    BP sweeps & \HypBudgetSweeps{} per shot & message-passing sweeps, summed over every BP call \\
    operation calls & \HypBudgetOpCalls{} per shot & calls of any DSL operation; the budget of last resort for generated loops \\
    candidates & \HypBudgetCandidates{} per shot & candidate corrections enumerated and tested \\
    column operations & \HypBudgetColumnOps{} per shot & GF(2) column operations spent in elimination (OSD, LSD) \\
    work cap & \HypWorkCap$\times$ P2-d & mean priced work per shot over one evaluation \\
    \bottomrule
  \end{tabular}
\end{table}

\paragraph{Paired adjudication.}
All decoders decode the same shots (common random numbers), so comparisons are paired: McNemar's test on the discordant counts $b$ (we succeed, they fail) and $c$ (the reverse), $z = (b-c)/\sqrt{b+c}$.
Dominance requires the test in \emph{both} directions, so that low power cannot be read as parity.
The oracles named above are configured independently of our own settings, not handed ours.

\paragraph{The bars and the verdicts.}
\emph{Primary}: Pareto-dominated by no panel member at its best known settings, and strictly dominating at least one.
\emph{Stretch}: more accurate than the reference P2-d at a cost lower by the margin of \OctAdjMargin$\times$.
\emph{Reach}: the accuracy of BP+OSD-CS at that cost.
A cost verdict must hold on both rulers, the work proxy and pinned-core time (Appendix~\ref{app:work}).
Overall, \OctAdjPrimaryProxy{} subjects met primary on the proxy and \OctAdjPrimaryTime{} on time, \OctAdjStretchTime{} met stretch on both (\OctAdjStretchWinners), and \OctAdjReachTime{} met reach.
Each decoder runs at its own budgets (the campaign's for the subjects, the published settings for the panel), and the reference saturates none of them.

\paragraph{The slate.}
Table~\ref{tab:slate} is the slate as read.
In all, \OctAdjSubjects{} subjects were frozen in writing before the read: the champion each run pinned, and the top-ranking archive finalists from the validation split; the table lists every one but one seeded finalist that we do not report, and the verdict counts above include it.
C13, a seeded finalist, is the most accurate subject (LER \OctAdjETwoLer{} at \OctAdjETwoCostXPTwoD$\times$ P2-d's proxy cost), but it raised an error on \OctAdjETwoRaisedShots{} of its timing shots, so its time ruler was refused and it meets primary on the proxy only; it is not reported in the body.

\paragraph{Time units.}
Time on this split is in P2-d units on one pinned core of an \OctAdjMachine{}, under two calibrations of one CPU model: the subjects under the campaign's own (P2-d at \OctAdjUnitNOne{} $\mu$s per shot, drift \OctAdjDrift), and the panel rows under the earlier read's (\OctAdjUnitPanel{} $\mu$s), in which the campaign's P2-d reads \OctAdjPTwoDInNOneUnits.
So a subject's time over a panel row's, such as C14's \OctAdjETwentySevenTimeXAc$\times$ BP+AC's, spans the two calibrations.
P5 here is our DSL port of Relay-BP at the authors' tuned point (LER \OctAdjPFiveLer{} at \OctAdjPFiveTimeXPTwoD{} P2-d units), not the authors' package; the body's Relay-BP rows are the package, timed on the benchmark workstation.

\begin{table}[ht]
  \centering
  \caption{%
    The slate on the \OctAdjShots{} shots of split \OctAdjSplit{}, sorted by LER.
    Proxy and time are in P2-d units (time on one pinned core); $z$ is McNemar, positive when the subject is right more often than the named panel member (P2-d the reference, P10 Ambiguity Clustering as published).
    Two further finalists make the same decisions as C10 and C06 on every shot and are not repeated.
  }
  \label{tab:slate}
  \footnotesize
  \setlength{\tabcolsep}{3pt}
  \begin{tabular}{llrrrrrl}
    \toprule
    subject & arm & LER & proxy & time & $z$ vs P2-d & $z$ vs P10 & bar \\
    \midrule
    C13 & seeded, wall-clock & \OctAdjETwoLer & \OctAdjETwoCostXPTwoD & refused & & & primary (proxy) \\
    C14 & seeded, proxy & \OctAdjETwentySevenLer & \OctAdjETwentySevenCostXPTwoD & \OctAdjETwentySevenTimeXPTwoD & \OctAdjETwentySevenZVsPTwoD & \OctAdjETwentySevenZVsPTen & primary \\
    C15 & seeded, proxy & \OctAdjETwentyFiveLer & \OctAdjETwentyFiveCostXPTwoD & \OctAdjETwentyFiveTimeXPTwoD & \OctAdjETwentyFiveZVsPTwoD & \OctAdjETwentyFiveZVsPTen & primary \\
    C10 & random, wall-clock & \OctAdjCTenLer & \OctAdjCTenCostXPTwoD & \OctAdjCTenTimeXPTwoD & \OctAdjCTenZVsPTwoD & \OctAdjCTenZVsPTen & primary \\
    C04 & random, proxy & \OctAdjCFourLer & \OctAdjCFourCostXPTwoD & \OctAdjCFourTimeXPTwoD & \OctAdjCFourZVsPTwoD & \OctAdjCFourZVsPTen & primary \\
    C08 & seeded, wall-clock & \OctAdjCEightLer & \OctAdjCEightCostXPTwoD & \OctAdjCEightTimeXPTwoD & \OctAdjCEightZVsPTwoD & \OctAdjCEightZVsPTen & stretch \\
    C03 & seeded, proxy & \OctAdjCThreeLer & \OctAdjCThreeCostXPTwoD & \OctAdjCThreeTimeXPTwoD & \OctAdjCThreeZVsPTwoD & \OctAdjCThreeZVsPTen & stretch \\
    C09 & seeded, wall-clock & \OctAdjCNineLer & \OctAdjCNineCostXPTwoD & \OctAdjCNineTimeXPTwoD & \OctAdjCNineZVsPTwoD & \OctAdjCNineZVsPTen & stretch \\
    C07 & seeded, wall-clock & \OctAdjCSevenLer & \OctAdjCSevenCostXPTwoD & \OctAdjCSevenTimeXPTwoD & \OctAdjCSevenZVsPTwoD & \OctAdjCSevenZVsPTen & stretch \\
    C06 & random, proxy & \OctAdjCSixLer & \OctAdjCSixCostXPTwoD & \OctAdjCSixTimeXPTwoD & \OctAdjCSixZVsPTwoD & \OctAdjCSixZVsPTen & primary \\
    C11 & random, wall-clock & \OctAdjCElevenLer & \OctAdjCElevenCostXPTwoD & \OctAdjCElevenTimeXPTwoD & \OctAdjCElevenZVsPTwoD & \OctAdjCElevenZVsPTen & --- \\
    C05 & random, proxy & \OctAdjCFiveLer & \OctAdjCFiveCostXPTwoD & \OctAdjCFiveTimeXPTwoD & \OctAdjCFiveZVsPTwoD & \OctAdjCFiveZVsPTen & --- \\
    C12 & random, wall-clock & \OctAdjCTwelveLer & \OctAdjCTwelveCostXPTwoD & \OctAdjCTwelveTimeXPTwoD & \OctAdjCTwelveZVsPTwoD & \OctAdjCTwelveZVsPTen & --- \\
    \midrule
    P5 Relay-BP (our port) & panel & \OctAdjPFiveLer & \OctAdjPFiveCostXPTwoD & \OctAdjPFiveTimeXPTwoD & & & \\
    P10 BP+AC, as published & panel & \OctAdjPTenLer & \OctAdjPTenCostXPTwoD & \OctAdjPTenTimeXPTwoD & & & \\
    P3 BP+OSD-CS, \ldpc{}'s point & panel & \OctAdjPThreeLer & \OctAdjPThreeCostXPTwoD & & & & \\
    P2 BP+OSD-0, \ldpc{}'s point & panel & \OctAdjPTwoLer & \OctAdjPTwoCostXPTwoD & & & & \\
    P2-d (the reference) & panel & \OctAdjPTwoDLer & \OctAdjPTwoDCostXPTwoD & \OctAdjPTwoDTimeXPTwoD & & & \\
    \bottomrule
  \end{tabular}
\end{table}

\section{The cost model}
\label{app:work}

Selection ranked on a \emph{work proxy}: a non-negative least-squares fit of per-shot wall time on one pinned core onto \CostCounters{} deterministic counters, frozen once fitted, in units of the reference P2-d's mean (\CostRef{} weighted units per shot).
Table~\ref{tab:weights} gives the counters and their frozen weights; a min-sum edge update costs \CostMinsumOverProdsum{} of a product-sum one.
Over \CostFitPrograms{} programs the fit's 95th-percentile relative error is \CostFitPNinetyFive{} in the fit and \CostFitHoldoutPNinetyFive{} held out ($R^2 = \CostFitRSq$), and its rank agreement with measured time is Kendall $\tau = \CostRankCFourTau$ (Spearman \CostRankCFourSpearman) on the fitting machine and $\tau = \CostRankMsukTau$ (Spearman \CostRankMsukSpearman) on the benchmark workstation.
A cost gap counts only past a margin of \CostSelectionMargin$\times$, set above the floor of \CostMarginFloor$\times$ that the fit's own error can resolve.

\begin{table}[ht]
  \centering
  \caption{The work proxy's counters and frozen weights, in units of one min-sum BP edge update.}
  \label{tab:weights}
  \small
  \begin{tabular}{lr}
    \toprule
    counter & weight \\
    \midrule
    BP edge updates, min-sum & \CostWeightEdgesMinsum \\
    BP edge updates, product-sum & \CostWeightEdgesProdsum \\
    GF(2) elimination words & \CostWeightGfTwoWords \\
    candidate-enumeration words & \CostWeightCandWords \\
    ranking and scan elements & \CostWeightScanElems \\
    DSL operation calls & \CostWeightOpCalls \\
    interpreter ticks & \CostWeightClastTicks \\
    \bottomrule
  \end{tabular}
\end{table}

\begin{figure}[ht]
  \centering
  \includegraphics[width=\textwidth]{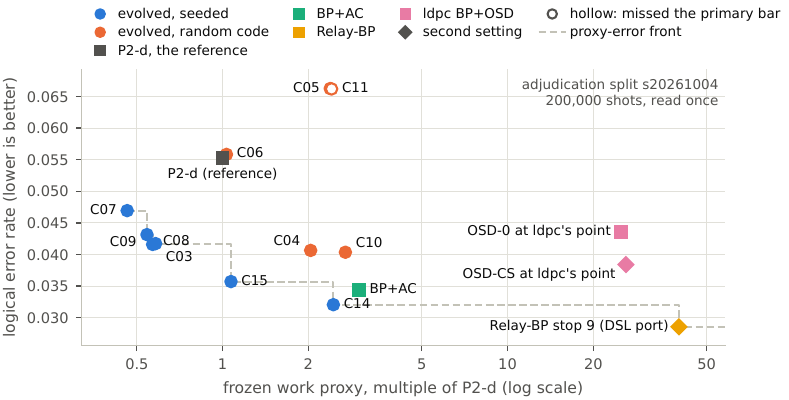}
  \caption{%
    Figure~\ref{fig:pareto} on the proxy ruler: logical error rate against the frozen work proxy in P2-d units (log scale), on the \OctAdjShots{} shots of split \OctAdjSplit{}.
    The \ldpc{}-named rows here are our DSL compositions at \ldpc{}'s tuned point, and Relay-BP at stop 9 is our DSL port, not the authors' package.
    Two finalists that duplicate C10 and C06 shot for shot are not drawn; C12 lies outside the frame.
    The front computed from the drawn rows is C07, C09, C08, C15, C14 and Relay-BP at stop 9 (our port).
  }
  \label{fig:pareto-proxy}
\end{figure}

\paragraph{Two rulers.}
Every adjudicated cost verdict is stated twice, in the proxy and in time on one pinned core, and must hold on both.
On every timed subject the two agree within a factor of \OctAdjProxyOverTimeMin–\OctAdjProxyOverTimeMax, and \OctAdjRefusals{} proxy verdicts were refused by the time ruler.
Figure~\ref{fig:pareto-proxy} is Figure~\ref{fig:pareto}'s companion on the proxy ruler, on the adjudication split.
The body quotes neither ruler: its times are clean ms per shot on the benchmark workstation, where the proxy's rank agreement with measured time is the $\tau = \CostRankMsukTau$ above.

\paragraph{The wall-clock arm.}
The wall-clock arm selected on measured time, calibrated per worker against P2-d: each run records \HypCalibrations{} calibrations, one per actor, and at \HypCpusPerActor{} CPUs per actor a calibration probe measured a median drift of \HypDriftMedianCpuTwo{}.
The run summaries record those calibrations, but they cannot show that each actor re-calibrated during the run.

\section{The twenty-two benchmark points}
\label{app:transfer}

The benchmark grid holds \GridPoints{} points: physical error rates $p = 0.001$ to $0.006$ at $n = 72$, code lengths $n = 90, 108, 144, 288$ at $p = 0.004$, and syndrome rounds 7 to 18 at $n = 72$, $p = 0.004$, each a circuit-level memory experiment in the $Z$ basis.
Every decoder keeps its exam-point settings at every point, because retuning one row per point would oblige us to retune every row, and the evolved rows, selected at one point, would gain the most from it.
Each point has its own shot count, from \GridShotsMax{} at $p = 0.001$ to \GridShotsMin{}, and times are clean ms per shot on the benchmark workstation except the \ldpc{} rows, timed on a second workstation of the same model and calibrated to the first.
Table~\ref{tab:fronts} counts the points at which each row is on the time–error front.
At $p = 0.001$, C03 makes \GridPOneCThreeErrors{} errors at \GridPOneCThreeMs{} ms and C08 \GridPOneCEightErrors{} at \GridPOneCEightMs{} ms, against BP+AC's \GridPOneAcErrors{} at \GridPOneAcMs{} ms, Relay-BP's \GridPOneRelayFiveErrors{} at \GridPOneRelayFiveMs{} ms (stop 5) and the reference's \GridPOnePTwoDErrors{} at \GridPOnePTwoDMs{} ms, on \GridPOneShots{} shots.
A clean pass of the Relay-BP package on the first \OctRelayShots{} shots of the adjudication split (\OctRelaySplit) reads \OctRelayFiveMs{} ms per shot at stop 5 and \OctRelayNineMs{} at stop 9, \OctRelayFiveTimeXPTwoD$\times$ and \OctRelayNineTimeXPTwoD$\times$ P2-d's \OctRelayUnitMs{} ms in the same session, consistent with the grid's \SepRelayFiveMs{} and \SepRelayNineMs{} ms at the exam-point.

\begin{table}[ht]
  \centering
  \caption{Points, of \GridPointsN{}, at which each row is on the time–error front over every benchmarked row. Fronts compare point estimates.}
  \label{tab:fronts}
  \small
  \begin{tabular}{lrlr}
    \toprule
    row & points & row & points \\
    \midrule
    C03 & \GridCThreeFrontN & BP+AC & \GridAcFrontN \\
    C09 & \GridCNineFrontN & Relay-BP, stop 9 & \GridRelayNineFrontN \\
    C04 & \GridCFourFrontN & Relay-BP, stop 5 & \GridRelayFiveFrontN \\
    C10 & \GridCTenFrontN & \ldpc{} BP+OSD-CS & \GridLdpcOsdCsFrontN \\
    C08 & \GridCEightFrontN & \ldpc{} BP+OSD-0 & \GridLdpcOsdZeroFrontN \\
    C05, C06, C07, C11, C12 & \GridCSevenFrontN & P2-d (the reference) & \GridPTwoDFrontN \\
    \bottomrule
  \end{tabular}
\end{table}

\section{Contribution 1 in detail: the damping knob and the damping pairs}
\label{app:damping}

\paragraph{What the factor of a hundred contains.}
On the \SepShots{} shots of the benchmark's exam-point (\SepSplit), \ldpc{}'s BP+OSD-0 at its tuned setting (product-sum BP, serial schedule) takes \SepLdpcOsdZeroMsExact{} ms per shot and errs on \SepLdpcOsdZeroLer{} of shots.
The same algorithm in the DSL with its BP stage re-parameterised —min-sum, flooding, damped— takes \SepPTwoDMs{} ms, about $1/\SepLdpcOsdZeroTimeXPTwoD$ of the time, and errs on \SepPTwoDLer{} ($z = \SepLdpcOsdZeroZVsPTwoD$ for the package).
That factor compares a package with the DSL, so it contains the difference between \ldpc{}'s C++ and our Rust engine as well as the parameter.

\paragraph{The parameter on its own.}
It is measured at the panel's front end: undamped flooding min-sum limit-cycles on this code, and on validation shots at equal iterations it errs on \KnobUndampedLerPct{} of shots against \KnobDampedLerPct{} damped —\KnobLerRatio$\times$ fewer errors at \KnobEdgesRatio$\times$ the BP edge updates.
\ldpc{} reproduces the failure when given these settings undamped (\KnobLiftedLerPct{} at \KnobLiftedMs{} ms per shot); its own stabiliser is the serial product-sum schedule, which is the expensive one.
So ``damping'' names the re-parameterisation that damping makes possible: the rule (product-sum to min-sum), the schedule (serial to flooding) and the damping itself, and for BP+OSD-0 the iteration count.

\paragraph{The pairs of Table~\ref{tab:damping}.}
Published front end: \OctDampPTwoSettings{}; damped front end: \OctDampPTwoDSettings{} (\OctDampPSevenDSettings{} for the LSD pairs).
Our sweep: \OctDampPThreeOursSettings{}; P3-d: \OctDampPThreeDSettings{}.
Time is clean ms per shot on the benchmark workstation (\OctDampCpu, one single-threaded process; \OctDampBlockShots{} timed shots a row after \OctDampBurnIn{} of burn-in; P2-d unit \OctDampUnitStart{} to \OctDampUnitEnd{} $\mu$s per shot, drift \OctDampDrift), and the LER is read on the \OctDampShots{} shots of split \OctDampSplit{}.
Every damped member is faster, by \OctDampOsdCsFrontSpeedup$\times$ to \OctDampOsdZeroSpeedup$\times$, and every damped member is less accurate, its LER \OctDampLerRatioMin–\OctDampLerRatioMax$\times$ the published member's ($z$ from \OctDampZMin{} to \OctDampZMax{} for the published member).
The speed is not fewer BP updates: the damped LSD members perform more edge updates per shot than the published ones and are still faster, because an edge of serial product-sum costs far more than an edge of flooding min-sum.
For BP+OSD-CS there are two damped variants: our own combination sweep, which also changes the back end, and its twin P3-d, the table's row, which keeps \ldpc{}'s OSD-CS and changes only the front end; the twin isolates the front end like the other three pairs, and it dominates our sweep, faster by \OctDampPThreeOursTimeXPThreeD$\times$ at a lower LER.

\section{The search-engine evidence}
\label{app:engine}

The decision to utilize a quality-diversity engine is grounded in a controlled, pre-campaign evaluation on this specific decoder problem. To isolate the impact of the search loop, we compared MAP-Elites against an otherwise identical generational engine, initializing both arms with a bit-identical generation zero.

\paragraph{Early Discovery (20 Generations):} At a constrained budget of twenty generations, the unseeded arm under MAP-Elites achieved a markedly higher held-out accuracy than the generational baseline.

\paragraph{Long-term Convergence (80 Generations):} We then extended this to an eighty-generation budget using a matched three-replicate control. At this scale, the performance margin narrowed, as the generational arm discovered its best solutions late in the run. A held-out rescore of the winners maintained the directional advantage of MAP-Elites, but the difference was not statistically resolved at only three replicates per arm.

The definitive claim supported by this control is that quality-diversity search finds strong decoders earlier; whether it finds strictly better ones at a full compute budget remains an open question. The main results presented in Section~\ref{sec:results} do not depend on this claim.

\end{document}